# ACOA - Chronological Analysis of the Exhibition of Artistic Works

Daniela Prado a), Armanda Rodrigues a), Nuno Correia a), Rita Macedo b) e c)  e Sofia Gomes b) e c)

a) NOVA LINCS and Informatics Department, NOVA School of Science and Technology, Universidade NOVA de Lisboa, Quinta da Torre, 2829-516 Caparica, Portugal
b) Department of Conservation and Restoration, NOVA School of Science and Technology, Universidade NOVA de Lisboa, Quinta da Torre, 2829-516 Caparica, Portugal
c) Instituto de História da Arte, NOVA School of Social Sciences and Humanities, Universidade Nova de Lisboa, Avenida Berna 26 C, 1069-061 Lisboa

## Abstract

We are currently prioritizing home activities, avoiding human contact, and carrying out external activities mostly by necessity. Therefore, and due to the loss of adhesion to cultural events on the part of the population, the cultural digital transformation process has been boosted, aiming to reach interested communities through digital media. The ACOA platform supports the organization of multiple sources of information related to creative processes behind complex artworks and their trajectories over time. This information is of great interest to conservators and curators, as well as to the general public, as it allows to document changes in the artwork, from the moment it was conceived by the artist, until its most recent exhibition. This platform houses a chronological evolution of the work, through the contextual dissemination of associated multimedia content. Works by the Portuguese artist Ana Vieira (1940-2016) were chosen as case studies for the implementation of the platform.

## 1. Introduction

When complex artworks are exhibited, particularly installations, almost all of the information transmitted to the observer is related to how the artist aims to present the work itself to its audience. There is, however, a whole process behind the artwork that led to its creation and exhibition. This information is potentially of great interest to conservators and curators, as well as to the general public, as it allows them to capture and document the trajectory of the artwork, from the moment it was conceived by the artist, until its most recent exhibition. Another factor of interest, which adds to knowledge of the biography of an artwork, is its reception by the artistic or general community, thus supporting the construction of the artist's reputation and her body of work.

Concurrently, the pandemic situation, arising from the spreading COVID-19 virus has led populations to prioritize their stay at home, to avoid human contact. External activities are thus being carried out by necessity and not for leisure, leading to the loss of adhesion to several sectors, among which the cultural sector. The artistic digital transformation process, which had already begun, was thus boosted, with the aim of reaching interested communities, through digital media.

The ACOA platform was created to fulfil the needs described above. The platform supports the process of chronologically documenting the evolution of artworks. This trajectory can include several moments, such as the creative process behind the construction of artworks, the different exhibitions held and reactions resulting from them. The transmission of all the desired information is achieved through the contextual dissemination of associated multimedia content, including image, video and audio. In addition, integrative representations in perspective (an even tri-dimensional) can be included, providing a spatial understanding of the works' display.

The platform is a generic software construction, which can be configured according to the user's needs (exhibitor, curator's public) to present the temporal and special trajectory of one or several artworks by an artist. It may also support the documentation of a complex exhibition of a set of artists, relating to an artistic movement or historical period. However, during its development, the artworks *Ensaio para uma Paisage*m (1997) and *Le Déjeuner sur L'Herbe*, (1977) by Portuguese artist Ana Vieira (1940-2016) were chosen as case-studies for the implementation.

The remainder of this paper is organized as follows: in section 2, the related work which supported some of the ideas for the development of this work is described. Section 3 presents ACOA and describes how the works of the artist Ana Vieira inspired its design and development. In section 4, the evaluation of the platform is presented. Finally, in section 5, some conclusions are drawn and perspectives for future developments are contemplated.

## 2. Related work

Early work into the design of ACOA focussed on evaluating online information tools supporting ideas which could contribute to the aim of the application to be developed. Early bibliographic search, on user centered design and the importance of using multimedia on web user interfaces provided support for the need to include diverse media available, while carefully considering the specificity of the types of materials used (Abras et al, 2004; Bradford, 2011; Mehta et al, 2012).

While considering the requirements of the tool, we concluded that the use of photographs of the artworks as well as the use of videos from the exhibition, would be advantageous, as they would allow the user to understand the complete work, as well as its components. However, the inclusion of a soundtrack, even when it is chronologically associated with the work and with the exhibition (as was the case of *Ensaio para uma Paisagem*) had to be treated differently. Although the music facilitates, to the user, the recreation of the exhibition environment, the automatic playing of the soundtrack is disadvised by the literature. Therefore, soundtrack playing was included in the tool as an option for the user.

Given the importance of the temporal context to this work, several online platforms, which include chronological analysis of materials were considered in the study, of which highlighted: RTP Museu Virtual (RTP, 2021), Google Arts & Culture (Google, 2021), On This Day (OnThisDay, 2021) and Evolution of the Web (EvolutionOfTheWeb, 2021). All these platforms were analysed with the aim of understanding their use of chronologies for temporal evolution, but only Google (2021) and RTP (2021) were studied from the point of view of the organization of art content, since they both provide cultural matter. The results of this analysis supported the choices made for the design of ACOA, including a horizontal chronology linking temporal events associated with the artwork, its exhibitions or events of the artist's life. The viewing of the materials as part of these relevant events, in a temporal continuity, should help users realize that there is a temporal evolution in the presented events.

## 3. ACOA

ACOA (Chronological Analysis of Artistic Works[1]) is an online platform that serves as a means of documenting works of art, presenting the chronological trajectory of its underlying artistic processes. The current configuration of the platform with two artworks by the Portuguese artist Ana Vieira, *Ensaio para uma paisagem* (1997) and *Le Déjeuner sur L'Herbe* (1977) was part of the initial motivation for the creation of the platform, a combination between an interest in promoting Ana Vieira's artistic processes and the existence of various materials (multimedia content) that could help sustain the dissemination of her oeuvre through a digital medium. The rising number of contemporary artist's legacies and the lack of information regarding their work led to the effort to creating this tool as a generic product.

The current configuration of ACOA is thus a tribute to Ana Vieira's work and career and will thus be described in the next section.

### 3.1.1. Current Configuration of ACOA

Ana Vieira's *Ensaio para uma* Paisagem (1997), was exhibited only once at the Natural History Museum in Lisbon, in Sala do Veado, in 1997. This installation is a good example to support the existence of the ACOA platform, since it was presented in a single exhibition and only those who watched it live could ever contemplate this work of art. It was the desire to give a new impulse to this work, exposing it in a digital medium where it can be viewed by many people, combined with the wish to convey, to this audience, the artist's entire creative process, that initially led to the effort to build this platform. The choice of using this artwork as the first implementation case-study in ACOA was thus simple. Additionally, the materials, documentation and testimonies available, related to this particular piece brought insight into Ana Vieira's artistic process. Thus, the chronology associated with *Ensaio para uma Paisagem* was divided into 3 phases: Conception, Exhibition and Post-Exhibition (Fig. 1). This structure allows the viewer to categorically capture the work, by addressing a temporal construction of the events.

---

[1] Análise Cronológica de Obras Artísticas

Because of the richness of testimonies, another level of depth has been considered into the main branch of the chronology, enabling subsections of content. The chronological phase of Conception includes the content associated with the time preceding the exposure of the installation. For a better organization of this phase, two subphases were indexed to the main branch of the chronology: Ideas, which shows the sketches made by the artist, and Materials, which described the materials that Ana Vieira used to create the components of the work. In the Exhibition chronological phase, the purpose was to present all the content captured during the exhibition, including photographic proofs and videos, as well as the sets of texts detailing the installation itself, as seen by the artist. Finally, the Post-Exhibition chronological phase includes the content that was created as a result of the exhibition of the work, such as press articles.

Figure 1 – Image of ACOA presentation of *Ensaio para uma Paisagem* © Arquivo Ana Vieira, © Ana Vieira Archive, courtesy of the family and Banco de Arte Contemporânea (BAC)

Ana Vieira's *Le Dejeuner sur L'Herbe* (1977) is a combination of Eduard Manet's original work with her personal proposal. In a dark room, the painting *Le Déjeuner sur*

*L'Herbe* by Eduard Manet is projected on top of a picnic towel laying on the floor with a variety of objects placed on it: four glasses, five plates, three bottles, one bowl, a palette with three brushes, a picnic basket in lintel and two oranges. Placing the painting projection at floor level forces the viewer to look down, an act of questioning the aesthetic experience of the painting. The decision to include this work into the platform came from the fact that it was exhibited several times, in different locations, which allowed us to make the best use of chronology, showing the potential of ACOA for transmitting the the artworks' trajectories. *Le Déjeuner sur L'Herbe* by Ana Viera (Fig. 2) is thus presented in 4 chronological phases: 1977, 1998, 2011 and 2017, coinciding with the years of the several exhibitions of the work and conveying continuity in time.

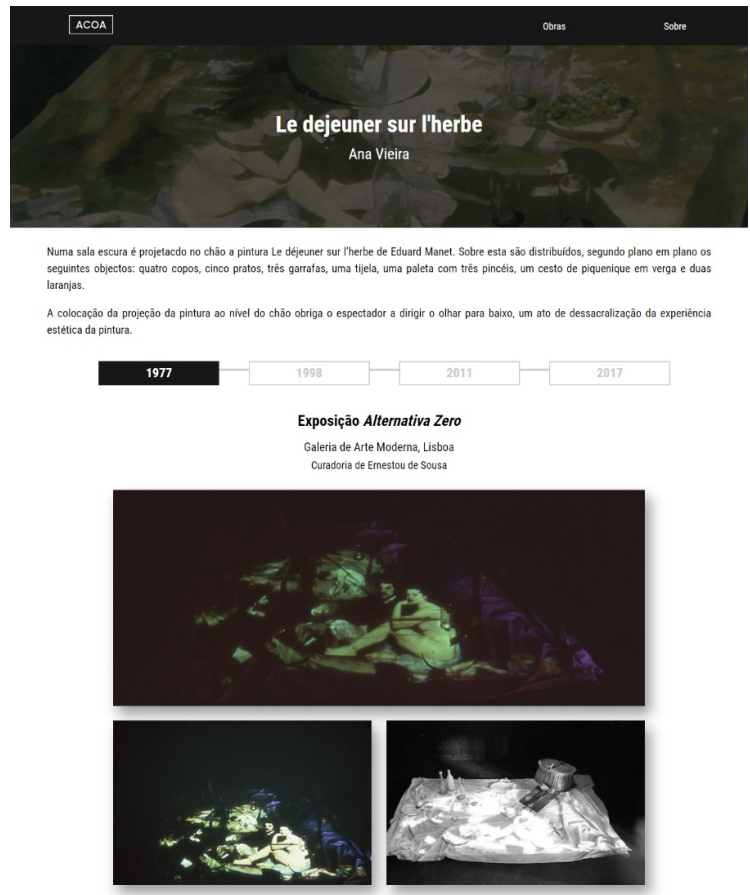

Figure 2 - Image of ACOA presentation of *Déjeuner sur L'Herbe* © Ana Vieira Archive, courtesy of the family and Banco de Arte Contemporânea (BAC)

The care given to both works by Ana Vieira in ACOA, helps to show its potential, not only by presenting structured content for different purposes (the artist creative process and the work's iterations over time), but also by showing how the chronology may either be qualitative and/or quantitative. Although the platform has currently only been tested with these two artworks, we aim to use it, in the future, to document the work of different artists. One of the aims of ACOA's design and development was genericity so that it can be easily configured for adding structure and materials. Therefore, in order to facilitate the entire process of inserting new artworks and associated multimedia materials, an editing tool was also developed, to be used only by administrative users. Consequently, the platform may be used by two profiles of users (administrative and public).

The common profile, to be accessed publicly by all interested users, allows access to all the content present in it. The online platform is divided into two sections: (1) The Works section and (2) the About section. The Works section is the main section of the platform and displays a list of all the works that make up the system. On the Works page, the user can have an immediate perspective of which and how many works he can obtain a more detailed knowledge about them (see Fig. 3). When viewing the list of works, the user can then choose to browse the works in that list, by clicking on the chosen artwork.

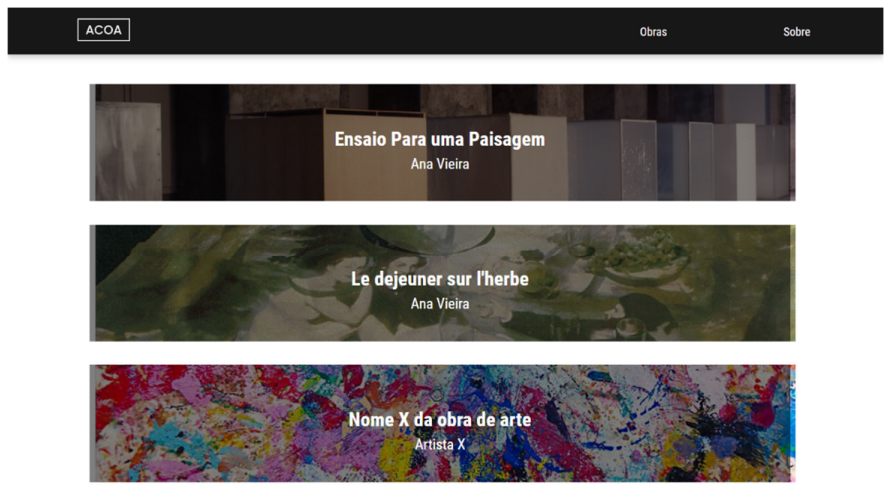

Figure 3 - Image of ACOA, images © Ana Vieira Archive, courtesy of the family and Banco de Arte Contemporânea (BAC)

From here, the user gets the perception of the selected work's trajectory through the viewing of the chronology that is associated with it and its relevant content. The webpages created for each chronological phase of an artwork have independent structures, since each work has its own chronology. The About section can be configured according to supported content of ACOA at a particular moment. Currently, the platform configuration is dedicated to Ana Vieira and, for this reason, the About section contains mainly biographic information about the artist.

The administrative profile can only be accessed by administrative users with specific access credentials. This section of the ACOA is thus hidden from ordinary users and focussed on the management of the available materials and documentation, to be made offered through the platform. This profile enables the expert user to add, edit and remove artworks and associated multimedia materials from the system. For each work, the following information is added to the platform: work title, artist (author), cover photo, the number of chronological stages/work's iterations, the name for each phase, image(s) to be associated with each chronological phase and descriptions of the chronological phase(s). Once all this information is added, the new work becomes accessible for viewing to the common user, as it will be featured in the Works list Section of the common part of the platform. The administrative user can also later edit the content associated with any available artwork. After selecting the work to edit, the user is taken to an editing interface, similar to the one used for adding new works, with pre-filled fields. Therefore, the user simply has to replace the value of the fields with the new information, where relevant. Additionally, it is also possible to change the number of phases/iterations of the chronology associated with the artwork. Finally, the administrative user can delete artworks from the platform. To do this, it is necessary to select the work to remove and finish the removal with a confirmation.

## 4. Platform Evaluation

In order to understand whether the ACOA platform was well structured, with a natural interface and whether it was responding to the needs of users, an evaluation methodology was developed and carried out with the help of different types of volunteer users. The testing methodology involved a diverse set of target users including volunteer non-specialists, computing students, conservation students and expert users, where the latter could primarily identify missing requirements in the developed product. The approach also involved the development of two types of tests: (1) a questionnaire and (2) a usability test based on the request to execute tasks in specific scenarios.

The questionnaire had the purpose of obtaining answers from the testers on all aspects of the platform, from their general opinions about it, to their opinions about the included features, and even information on whether the chronology helped to convey a message of temporal trajectory. In order to obtain more tangible results, the SUS (System Usability Scale) questionnaire (Sauro, 2011) was incorporated into the evaluation methodology, with the aim of generally evaluating the usability of the platform. The results of the evaluation were very positive, with measured values above average, such as the total time of execution of the scenario, the execution time of each task, user feedback about the ease of carrying out the scenario and the tasks, and the number of aids (interventions) needed by the users to finalize the scenario/task. Generic opinions collected from the testers were also positive.

## 5. Conclusions and future work

The current implementation of the ACOA platform is a prototype. However, it manages to show the potential it has and how far it can go. Despite currently presenting documentation with only two works by a single artist, the existence of an administrative profile enables its installation and configuration in different artistic contexts and settings, involving artworks by any artist. However, there is room for evolution of the current implementation, as future changes have been identified to be made in different parts of the platform.

Currently, several artworks can be added to the platform, each one including a particular set of chronological phases. Additional configuration potential is needed on the current product for improving detail on the definition of these phases and on the relationship between the information available for each artwork and the data conveyed in the About section. It is our intention to prepare the structure of ACOA to enable content on several artists which may be associated, in the platform, not only chronologically but also thematically, allowing the horizontal detailed structure to contemplate collectives of artists or exhibits. These changes will directly affect the interface design of the solution, which will have to be adapted to allow a better organization of the documented works, through categorization, and allowing the existence of filtering in the results presented to the user. The presence of categories will help to improve the user experience and, therefore, their navigation on the platform. Regarding the code, developed so far, it is expected that a reanalysis will be carried out, in order to detect possible improvements at the structural level. We would also like to contemplate the platform's capacity to handle tri-dimensional data, such as 3D images and models, captured in future exhibits, as well as the visualization of the temporal evolution of these structures.

**ACKNOWLEDGEMENTS**

This work is funded by FCT/MCTES NOVA LINCS UIDB/04516/2020.